\documentclass[aps,prb,twocolumn,superscriptaddress,showpacs,floatfix]{revtex4}
\usepackage{graphicx}
\usepackage{amsmath,amssymb}
\begin{document}
\title{Universality in adsorbate ordering on nanotube surfaces}
\author{V. I. Tokar}
\affiliation{IPCMS, Universit{\'e} de Strasbourg--CNRS, UMR 7504, 
23 rue du Loess, F-67034 Strasbourg, France}
\affiliation{Institute of Magnetism, NAS and MES of Ukraine,
36-b Vernadsky Boulevard, 03142 Kiev-142, Ukraine}
\author{H. Dreyss\'e}
\affiliation{IPCMS, Universit{\'e} de Strasbourg--CNRS, UMR 7504, 
23 rue du Loess, F-67034 Strasbourg, France}
\date{\today}
\begin{abstract}
Numerically efficient transfer matrix technique for studying statistics
of coherent adsorbates on small nanotubes has been developed.  In the
framework of a realistic microscopic model fitted to the data of {\em
ab initio} calculations taken from literature sources, the ordering of
potassium adsorbate on (6,0) single-walled carbon nanotube has been
studied.  Special attention has been payed to the phase transition-like
abrupt changes seen in the adsorption isotherms at low temperature.  It
has been found that the behavior during the transitions conforms with
the universality hypothesis of the theory of critical phenomena and is
qualitatively the same as in the one dimensional Ising model.
Quantitatively the critical behavior can be fully described by two
parameters.  Their qualitative connection with the properties of
interphase boundaries is suggested but further research is needed to
develop a quantitative theory.
\end{abstract}
\maketitle
\section{Introduction}
A considerable part of the ongoing research on adsorption in carbon
nanostructures is driven by the problem of hydrogen storage at ambient
conditions.\cite{nature97,Bernard07,Bianco09,review09}  In particular,
the metallic adsorbates are expected to considerably enhance the
hydrogen uptake\cite{Bernard07,cntTiH2,Kcnt60,CaH209,review09} because
the storage capacity of purely carbon structures is insufficient from
a practical point of view.\cite{Bianco09}  The adsorption of
gases\cite{scienceCNT2,N2_cnt07,gases02,cnt_ground_state07} allows,
{\em inter alia}, to gain deeper insight into the dependence of sorption on
various characteristics of adsorbate molecules, such as their
size.\cite{cnt_ground_state07}

From the storage perspective, the most promising among carbon
nanostructures are the single-walled nanotubes (SWNTs) because of their
large surface to weight ratio.\cite{Bernard07,Bianco09}  Since the
storage capacity is defined mainly by the adsorbing
surface,\cite{Bianco09} theoretical studies of the adsorption for
simplicity are often performed on individual
SWNTs.\cite{cntTiH2,Kcnt60,HonCNT,cnt_ground_state07,CaH209}  The
hydrogen uptake predicted in such studies is sometimes very
high\cite{cntTiH2,HonCNT,CaH209} but their significance for the storage
is not clear because the calculations are usually made for periodic
structures at zero temperature with only crude estimates of temperature
effects sometimes being made.\cite{cntTiH2,HonCNT}

Temperature effects, however, may strongly influence predictions based
on zero-temperature calculations.  For example, at finite temperatures
the ordered structures cannot exist in one-dimensional systems in the
thermodynamic limit.\cite{ll}  Instead, if temperature is sufficiently
low, a disordered state is formed with extended local order
corresponding to the $T=0$ K ordered structure.  From continuity
considerations it is reasonable to assume that at sufficiently low
temperature this quasi-ordered structure should be as good a hydrogen
absorber as the zero-temperature one.  Thus, from the storage point of
view the question is how large are the temperatures at which the
zero-temperature predictions can still be relied upon.  In the closely
related problem of adsorption on the two-dimensional (2D) surface this
question can be answered with the help of phase diagrams where ordered
and disordered phases are separated by well defined
boundaries.\cite{binderTM82,O/W}

The aim of the present paper is to adopt the techniques of Refs.\
\onlinecite{binderTM82,O/W} where adsorption of hydrogen and oxygen on
2D surfaces were investigated to the case of adsorption on individual
SWNTs and to find out what can be said about the quasi-ordered
structures in the absence of well defined finite-temperature phase
boundaries.  To implement this approach, some assumptions and
approximations need be made which are usually specific to the type of
the adsorbate under consideration.  For concreteness, we will discuss
them using as an example the potassium deposit on (6,0) zigzag SWNT.
This choice was motivated mainly by the fact that this system was
studied with an {\em ab initio} technique in Ref.\ \onlinecite{Kcnt60}
where the ground state energies of six periodic structures were
calculated.  Such information is necessary for the implementation of
the cluster expansion method (CEM)\cite{ECI,ducastelle,O/W,CEreview}
which allows one to derive an effective lattice gas Hamiltonian to use
in the solution of statistical problems.  In connection with
Refs.\ \onlinecite{ECI,ducastelle,CEreview} which deal with binary
alloys it is pertinent to point out that the lattice gas model is
formally equivalent to the binary alloy which allows for the use of
techniques developed in the alloy theory to coherent surface
adsorbates.  It should also be noted that in non-metallic systems the
CEM can be developed on the basis of the energies calculated with the
use of model potentials.\cite{cnt_ground_state07}  Furthermore, the
effective Hamiltonian can be derived via fit to experimental
data.\cite{binderTM82,O/W}

An important assumption made in the application of CEM to surface
structures is that the adsorbate is in registry with the substrate
lattice.  In reality, however, this depends on the relative strength of
interactions of adsorbate atoms between themselves and with the
substrate.  If the latter interaction is weak (which is the case in the
system under consideration\cite{alkali_review05}) the coherence with
the substrate in sufficiently dense structures can be
lost.\cite{referee1PRB}

Another difficulty is due to the substantial coverage-dependent charge
transfer which strongly influences the interactions of adsorbate atoms
with the substrate and with each other (see detailed discussion for the
potassium deposits on graphite, graphene, and on SWNTs in
Refs.\ \onlinecite{alkali_review05,Vasiliev}).  In principle, charge
transfer should be adequately accounted for in the {\em ab initio}
calculations.\cite{Kcnt60}  But the system under consideration is
apparently rather singular because the adsorption energy of potassium
atom on the graphen varies in different calculations from 0.44 to 2.0
eV.\cite{Vasiliev}  On that scale our neglect of phonons whose
characteristic Debye energy is usually an order of magnitude smaller
looks justified.  The question remains on the importance of the
entropic contribution at finite temperature due to the atomic
vibrations.  In the case of alloys this problem was reviewed in
Ref.\ \onlinecite{rmp2} where it was concluded that in the majority of
cases the classical approximation should be sufficient.  At least, this
is justifiable in the case of atoms with large atomic mass like
potassium.  The classical statistical averaging in the harmonic
approximation on which the phonon theory is based reduces to the
Gaussian integration over atomic coordinates which can be performed
exactly.  As explained in Appendix B of Ref.\ \onlinecite{PRB}, the
energy part of the configuration free energy thus obtained coincides
with the energy minimum at zero temperature.  Thus, the equilibrium
atomic positions obtained in the {\em ab initio}
calculations\cite{Kcnt60} at zero temperature provide the necessary
average energy while the entropic part can be unified with the
interaction part as an effective temperature-dependent contribution
into the pair interatomic interaction.\cite{rmp2,PRB} Because in the
present study we are going to consider temperatures which are small in
comparison with the pair interactions, we will neglect this
contribution in our calculations.

Finally, when deposited atoms or molecules are very light (He and H$_2$
being the most important examples), quantum corrections became
important at low temperatures and quantum treatment is
preferable.\cite{quantum_effects,He_cnt}  In Ref.\
\onlinecite{quantum_effects}, however, it was shown for the hydrogen
molecules adsorbed in the nanotube bundles that a fully classical
regime sets in already at temperatures above 60 K.  Because for the
storage purposes the temperatures below the liquid nitrogen boiling
point (77 K) are of little interest,\cite{review09} the quantum
corrections can be neglected in the studies oriented on storage
applications.

Thus, the main difficulties in the statistical description of the
adlayers on SWNTs are due to the incommensurate structures and the poor
accuracy of the interaction parameters.  The accuracy can be improved
either with the use of a better {\em ab initio} approach or by a direct
fit to experimental data.\cite{binderTM82}  The loss of coherence with
the substrate is a more serious problem because the lattice gas
formalism usually requires a regular lattice to exist (see, however,
Sec.\ 5.3 of Ref.\ \onlinecite{CEreview}).  In adsorbates this usually
restricts the coverages at which the system retains its coherence to
low values $\lesssim0.5$.\cite{alkali_review05,scienceCNT2}

But on the other hand, as is well known (see, e.\ g.,
Ref.\ \onlinecite{ducastelle}), the lattice gas model is equivalent to
the Ising model which is famous for being capable of describing such
disparate critical phenomena as the magnetic ordering in uniaxial
magnets and the liquid-gas phase
transition.\cite{wilson_ea72,1Dliquid86,2Dliquid92}  This similarity
between the critical phenomena has been conceptualized in the
universality hypothesis which has been amply confirmed by both
experimental data and theoretical calculations in 2- and 3D
systems.\cite{universality72,stanleyRMP} Hopefully it will work equally
well in 1D systems\cite{D1-D4phaseD,nelson_fisher} which may allow for
the extension of our results to the incommensurate cases as well.

In view of the many approximations and assumptions which need be
accepted in order to implement the statistical approach to the
adsorption on SWNTs, in the present paper we will focus on the low
temperature regions in the vicinity of critical points where the
universal behavior sets in and where even significant inaccuracies in
the microscopic description in most cases may be irrelevant.

In the next section we will explain the universality hypothesis for 
one-dimensional systems belonging to the Ising universality class;  in
Sec.\ \ref{model} the effective Hamiltonian will be derived in the
framework of the CEM; in Sec.\ref{solution} the partition function will
be calculated with the use of numerically efficient transfer matrix
(TM) technique and in Sec.\ \ref{discussion} we present our
conclusions.
\section{Universality in 1D}
Universality hypothesis constitutes one of pillars of the modern
theory of critical
phenomena.\cite{universality72,D1-D4phaseD,stanleyRMP} It
states that the singular part of the equations of state of all systems
belonging to the same universality class has the same functional form
in the vicinity of the critical point.  Unique for each system are only
two constant parameters which define the scales of variation of the
(dimensionless) external field
\begin{equation} 
\label{ L}
L = h/k_BT
\end{equation} 
and of the reduced temperature 
\begin{equation} 
\label{ t}
t\equiv(T-T_c)/T_c,
\end{equation} 
where $T_c$ is the critical temperature.  Thus, one may simplify the
task of predicting the behavior of a system in the critical region by
solving the simplest model belonging to the universality class of
interest.  The two parameters can be either found in independent
calculations or derived from experimental data.  For the general
discussion of the universality we refer the interested reader to the
vast literature on the
subject\cite{universality72,D1-D4phaseD,stanleyRMP} while below we will
consider only the Ising universality class in
1D.\cite{prange74,nelson_fisher,priest75} The peculiarity of this case
is that there is no finite-temperature phase transitions in 1D.
Therefore, the approach to universality based on scaling variables
(\ref{ L}) and (\ref{ t}) cannot be applied straightforwardly because
$T_c=0$ and the scaling variable $t$ is undefined.

A solution to this problem was found in
Ref.\ \onlinecite{nelson_fisher}.  It was noted that instead of the
scaling parameter (\ref{ t}) the correlation length $\xi$ can be used
due to the relation
\begin{equation} 
\label{ xi}
t\sim\xi^{-1/\nu},
\end{equation} 
where $\nu$ is the critical index which defines the divergence of the
correlation length as $\xi\sim t^{-\nu}$.  Taking into account that in
1D all critical indices are known exactly, on the basis of Eqs.\
(3.40)--(3.41) of Ref.\ \onlinecite{nelson_fisher} the equation of
state in the scaling region can be written as
\begin{equation}
\label{ eq0}
M\approx W(L\xi/2),
\end{equation}
where $M$ is the magnetization normalized as 
\begin{equation} 
\label{ M}
M(\pm\infty)=\pm1
\end{equation} 
and $W$ is the scaling function. The latter is universal for all
systems belonging to the Ising universality class in 1D except for two
constant factors: one factor multiplying $W$ thus changing the range of
variation of $M$ in Eq.\ (\ref{ M}) and another one before its
argument.  We used this arbitrariness in Eq.\ (\ref{ eq0}) by dividing
the argument by two in comparison with
Ref.\ \onlinecite{nelson_fisher}.  This definition is more appropriate
to our purposes and according to Eq.\ (\ref{ xi}) it does not change the
scaling relations which are invariant under rescalings.

The above formalism can be easily refashioned to describe the lattice 
gas model via the equivalence transformation
\begin{equation}
\label{ s2n}
\sigma_i=2n_i-1,
\end{equation}
where $\sigma_i=\pm1$ is the Ising spin on site $i$ and $n_i=0,1$ the
corresponding occupation number.  From this identity it follows that
\begin{equation} 
\label{ H-mu}
h\sigma_i = \mu n_i -\mu/2,
\end{equation} 
where $\mu = 2h$.

The coverage is defined as the lattice gas density
\begin{equation} 
\label{ rho}
\rho=\langle n_i\rangle,
\end{equation} 
where the angular brackets denote statistical averaging and the
dependence of $\rho$ on the lattice site is absent because the system
is assumed to be homogeneous and the spontaneous symmetry breaking is
absent in 1D.\cite{ll}

According to Eqs.\ (\ref{ s2n}) and (\ref{ H-mu}) Eq.\ (\ref{ eq0}) 
takes the form
\begin{equation}
\label{ eq1}
(\rho-\rho_c)/(\Delta\rho/2) \approx W[(\mu-\mu_c)\xi/k_BT],
\end{equation}
where $\mu_c$ is the chemical potential at the critical point and
$\Delta\rho=\rho_{+}-\rho_{-}$ is the total change of the density
during the transition.  Because at the critical point the critical
density $\rho_c\approx(\rho_{+}+\rho_{-})/2$, the left hand side of
Eq.\ (\ref{ eq1}) varies in the same range as in Eq.\ (\ref{ M}).  We
note that to achieve this we had to divide $\rho-\rho_c$ on the right
hand side of Eq.\ (\ref{ eq1}) by $\Delta\rho/2$, not by $\rho_c$ as
suggested in Ref.\ \onlinecite{universality72}.  In this way we fix
one of the arbitrary scale factors in our problem.

Thus, according to the universality principle the behavior of the
system near any critical point can be describe with the use of only two
constant scale factors provided the universal function $W$ is known.
The latter can be calculated for the simplest possible model, the Ising
model with the fist neighbor interactions being the most obvious
choice.
\subsection{\label{1D}1D lattice gas model} 
Exact solutions of the 1D Ising model can be found in many places, for
example, in Eq.\ (3.39) of Ref.\ \onlinecite{nelson_fisher} or in
Ref.\ \onlinecite{priest75}.  But below for completeness we present the
solution of the equivalent lattice gas model with the use of a variant
of the nonsymmetric transfer matrix
technique\cite{screwed_cilinder,screwed_cilinder2} which in
Sec.\ \ref{solution} will be generalized to the case of nanotubes.

In the process of adsorption the number of atoms on the surface is
governed by the chemical potential $\mu$ which may be controlled by the
gas pressure  if the adsorption from gaseous phase takes place (see,
e.\ g., Ref.\ \onlinecite{HonCNT}) or by the concentration of the
adsorbate in the solution in the case of adsorption from a liquid.
Therefore, the natural choice is the grand ensemble formalism
with the partition function
\begin{equation} 
\label{ Xi_def}
\Xi = \underset{n_i=0,1}{\text{Tr}}\,e^{-\beta H},
\end{equation} 
where $\beta = 1/k_BT$ and $H$ the Hamiltonian; for brevity the term
with the chemical potential $-\mu\sum_in_i$ is considered to be
included into $H$.\cite{binderTM82}  With the use of Eq.\ (\ref{
Xi_def}) the coverage can be found as [cf.\ Eq.\ (\ref{ rho})]
\begin{equation} 
\label{ rho0}
\rho=(\beta N)^{-1}d\ln\Xi/d\mu,
\end{equation} 
where $N$ is the number of deposition sites. In the case of 1D
lattice gas with only nearest neighbor interaction $V_1$
(which we assume to be attractive) Eq.\ (\ref{ Xi_def}) can be written
as
\begin{equation} 
\label{ Xi1D}
\Xi_{1D} = \underset{n_j=0,1}{\text{Tr}}\,
e^{\beta\mu n_N}\prod_{i=1}^{N-1}\exp(\beta\mu n_i -\beta V_1n_in_{i+1}).
\end{equation} 
We assume free boundary conditions corresponding to nanotubes with open
ends.  According to Eq.\ (\ref{ Xi1D}), $\Xi_{1D}$ can be calculated
via the $N-1$-st power of the transfer matrix
\begin{equation} 
\label{ T}
\hat{T}=\left(
\begin{array}{cc}
1 & 1\\
e^{\beta\mu}& e^{\beta(\mu-V_1)}
\end{array}
\right).
\end{equation} 
In the thermodynamic limit the reduced free energy Eq.\ (\ref{
phi_def}) of the 1D lattice gas is
\begin{equation} 
\label{ phi}
\phi_{1D} = -\beta^{-1}\ln\lambda_+,
\end{equation} 
where $\lambda_+$ is the largest eigenvalue of $\hat{T}$.  The
logarithm in Eq.\ (\ref{ phi}) can be cast into the form
\begin{equation} 
\label{ log_lambda}
\ln\lambda_+ =  \delta + \ln(\cosh\delta+\sqrt{\sinh^2\delta 
+e^{\beta V_1}}),
\end{equation} 
where 
\begin{equation} 
\label{ delta}
\delta=\beta(\mu - V_1)/2\equiv\beta(\mu - \mu_c)/2.
\end{equation} 
The coverage can be found as 
\begin{equation} 
\label{ Dlog_l}
\rho=\beta^{-1}d\ln\lambda_+/d\mu = \frac{1}{2} 
+ \frac{1}{2}\frac{\sinh\delta}
{\sqrt{\sinh^2\delta +e^{\beta V_1}}}.
\end{equation} 
As is easy to see, at low temperature $e^{\beta V_1}\to0$ and
$\rho(\delta)$ tends to the $\theta$-function.  In other words, all
variation of  $\rho(\delta)$ is restricted to a narrow interval of
$\mu\approx\mu_c$.  That is why this region is so important.  The
interaction potential and the temperature may vary in very broad ranges
but the values $\rho=0$ to the left of the interval and $\rho=1$ to the
right of it will remain the same. In other words, these ranges of
variation of $\mu$ do not provide much useful information on the
microscopics of the system.  In contrast, in the vicinity of $\mu_c$
the slope
\begin{equation} 
\label{ chi}
\frac{d\rho}{d\mu} = \frac{\beta}{4}  
\frac{e^{\beta V_1}\cosh\delta}{(\sinh^2\delta+e^{\beta V_1})^{3/2}}  
\end{equation} 
will vary very strongly with temperature, with $V_1$, with $\mu$, etc.,
so all quantities of interest are most easily measured near this
quasi-transition point.

At $\mu=\mu_c$ $\rho$ in Eq.\ (\ref{ Dlog_l}) is equal to 0.5.  This
means that both phases---$\rho=0$ and $\rho=1$---are present in the
system in equal proportion. At low temperature according to
Eqs.\ (\ref{ drhodm}) and (\ref{ chi}) the atoms are correlated at long
distances so the system looks as an intermittent mixture of the pure
phases separated by interphase boundaries (IPBs).  In the model under
consideration the boundary energy at $T=0$ is easy to calculate.  In
the Ising spin representation Eqs.\ (\ref{ s2n})--(\ref{ H-mu}) the 
spin-spin interaction at $h=0$ ($\rho=0.5$) is
\begin{equation} 
\label{ spin_repr}
(V_1/4)\sum_i\sigma_i\sigma_{i+1}=V_1\sum_in_in_{i+1}-\mu N/2.
\end{equation} 
At zero temperature the IPB will separate the region of spins up 
from the spin down region.  According to Eq.\ (\ref{ spin_repr}), in
comparison with the ordered system the energy cost is
\begin{equation} 
\label{ E_b}
E_b=|V_1|/2.
\end{equation} 
Furthermore, in a system with
free boundaries there are two equally probable possibilities for an
IPB:  $\uparrow\downarrow$ and $\downarrow\uparrow$.  Thus, there is
the entropy $k_B\ln2$ associated with the IPB.

In general case we may introduce the free energy of the IPB as
\begin{equation} 
\label{ G_b}
G_b=H_b-TS_b,
\end{equation} 
where $H_b$ is the enthalpy of the boundary creation and $S_b$ its
entropy.  The IPBs break long range correlations between different
parts of the system.  Therefore, the correlations extend at the
distances which are inversely proportional to the IPB concentration.
The latter can be estimated as\cite{ll}
\begin{equation} 
\label{ c_b}
c_b = e^{-\beta G_b}.
\end{equation} 
As can be seen from the explicit 1D solution above, $c_b$ also defines
the width of the region around $\mu_c$ where the fast change of the
adsorption isotherm takes place.  This can be visualized with the use
of the variable $x$ introduced as
\begin{equation} 
\label{ x}
\beta(\mu - \mu_c)=c_bx.
\end{equation} 
With the use of this variable one can establish on the basus of
Eq.\ (\ref{ Dlog_l}) the explicit form of the universal function in
Eqs.\ (\ref{ eq0}) and (\ref{ eq1}) as
\begin{equation} 
\label{ W}
W(x)\approx\frac{x}{\sqrt{1+x^2}}.
\end{equation} 
We note, that $W(\pm\infty)=\pm1$, as necessary.  From practical point
of view, Eqs.\ (\ref{ eq1}) and (\ref{ Dlog_l}) are not very convenient
for universality checks because both sides of these equations turn into
zero at the critical point $\rho=\rho_c$.  This means that in the data
measured or calculated with finite precision on a non-singular
background the universal behavior can be obscured by the errors.  The
singular behavior can be considerably enhanced by differentiation with
respect to the gas pressure (see Fig.\ 1 in
Ref.\ \onlinecite{mursic96}) or with respect to the variable $x$:
\begin{equation} 
\label{ scaling}
\frac{d\rho}{d x} = \frac{1}{2(1+x^2)^{3/2}}.
\end{equation} 
As will be shown in Sec.\ \ref{solution}, this expression indeed
describes the isothermal compressibility of coherent deposits on SWNTs
near the steps of the adsorption isotherms.
\section{\label{model}The model}
The configuration of a coherent deposit consisting of identical atoms
or molecules in the submonolayer coverage regime can be fully
characterized by the occupation numbers $n_i=0,1$ of the deposition
sites $i=1,N$.  The configuration energy of the deposit can be expanded
into an infinite series of effective cluster interactions (ECIs)
as\cite{ECI,ducastelle,O/W}
\begin{eqnarray} 
\label{ H}
H^{(N)}=&&(E_{ads}-\mu)\sum_{i}n_i+\sum_{i>j}V_{ij} n_in_j\nonumber\\
&&+\sum_{i>j>k}V_{ijk}^{(3)} n_in_jn_k+\dots,
\end{eqnarray} 
where we assumed that the system is homogeneous so the adsorption
energy $E_{ads}$ is the same at each site; also, as in Eq.\ (\ref{
Xi_def}), we included into the Hamiltonian the chemical potential $\mu$
to control the coverage.  To find the interaction parameters in
Eq.\ (\ref{ H}), one needs, according to established
methodology,\cite{ECI,ducastelle,O/W} to compute the energies of a
sufficiently large number of different adsorbate structures and then fit
these energies to the lattice gas Hamiltonian (\ref{ H}) with
sufficient number of ECIs.  The energies are usually calculated {\em ab
initio} but model calculations present viable
alternative.\cite{ECI,ducastelle,O/W}  Yet another possibility is to
adjust the interactions to the experimental data (see the discussion
and the bibliography in 
Ref.\ \onlinecite{O/W}).\cite{ground_states03,ArV_g,V_gXe08,KrV_g08}

In our calculations below we consider the adsorption of potassium on
the surface of the zig-zag (6,0) carbon nanotube studied in
Ref.\ \onlinecite{Kcnt60} where the energies of six ordered structures
were calculated within an {\em ab initio} approach.  We remind that
potassium and other metal deposits are directly related to the problem
of hydrogen storage.\cite{science99,cntTiH2,Kcnt60,CaH209,review09}

\begin{figure}
\begin{center}
\includegraphics[viewport = 120 620 312 740, scale = 1]{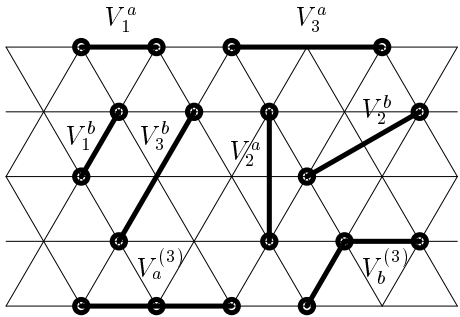}
\end{center}
\caption{\label{fig1}Effective cluster interactions used to fit
Hamiltonian in Eq.\ (\ref{ H}) to the {\em ab initio} data of Ref.\
\onlinecite{Kcnt60}.  The triangular lattice of deposition sites is
shown (the deposition site is defined as the center of the carbon
hexagon, see Fig.\ \ref{fig2}). The tube axis is directed vertically.
The leftmost and the rightmost sites in the odd rows on the drawing
represent the same site on the tube.}
\end{figure}

\begin{figure}
\begin{center}
\includegraphics[viewport = 15 750 612 810, scale = .7]{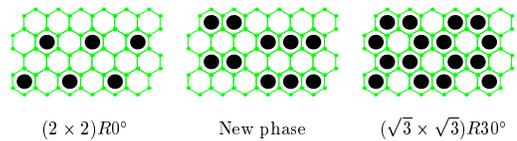}
\end{center}
\caption{\label{fig2}The ground state structures on the surface of
(6,0) SWNT with fillings 1/4, 5/12, and 2/3 found in Monte Carlo
simulated annealing described in the text.  The small dots correspond
to the carbon atoms and the large dots the potassium atoms.  The ranges
of stability of these structures with the change of the chemical
potential are shown on Fig.\ \ref{fig3}.}
\end{figure}

From the structures considered in Ref.\ \onlinecite{Kcnt60} one can
deduce that at least third neighbor pair interactions need be included
into the Hamiltonian (\ref{ H}) (see their Fig.\ 1 and our
Figs.\ \ref{fig1}--\ref{fig2}).  Because of the tube anisotropy, the
number of pair interactions is, in fact equal to six, as shown in
Fig.\ \ref{fig1}.  This is equal to the number of energy values we have
at our disposition which is insufficient even to fit the pair
interactions because we need also to determine the adsorption energy
$E_{ads}$.
\begin{figure}
\begin{center}
\includegraphics[viewport = 60 240 512 525, scale = 0.45]{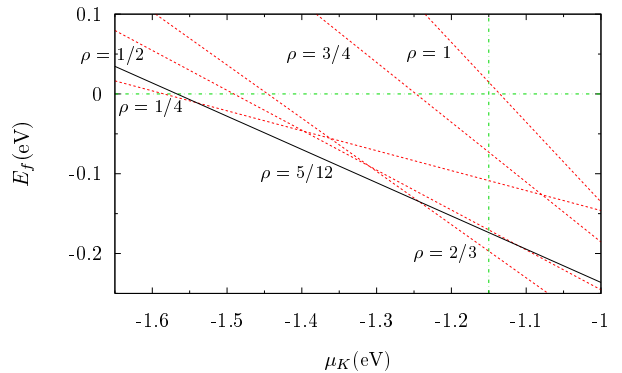}
\end{center}
\caption{\label{fig3}Zero temperature phase diagram of potassium
adsorbed on the surface of (6,0) carbon nanotube derived on the basis
of Hamiltonian in Eq.\ (\ref{ H}) fitted to the data of Ref.\
\onlinecite{Kcnt60}.  The dotted lines represent the fit to some of the
structures found in that reference.  The solid line corresponds to the
new phase shown in Fig.\ \ref{fig2}. $E_f$ is the energy of formation
of the adsorbed structure and $\mu_K$ the chemical potential of
potassium.  The vertical line corresponds to bulk
potassium.\cite{Kcnt60}}
\end{figure}

To overcome this difficulty we, following Ref.\ \onlinecite{morseK/Cu}, 
assume that the pair interactions between the potassium atoms can be 
approximately described by the Morse potential
\begin{equation} 
\label{ morse}
V(r_{ij}) = \epsilon(e^{-2a(r_{ij}-r_0)}-2e^{-a(r_{ij}-r_0)})
\end{equation} 
which depends on three parameters $\epsilon$, $a$, and $r_0$.
Taking into account $E_{ads}$, we are left with the possibility to
adjust two more parameters.  This turned out to be indispensable
because the data of Ref.\ \onlinecite{Kcnt60} could not be fitted to
the Hamiltonian containing only pair interactions.  This can be shown
by expressing the energies of the structures in terms of (unknown) pair
interactions and then establishing exact relations between some of the
energies by excluding the pair interactions from the expressions.  The
sum rules obtained in this way are strongly violated by the data of
Ref.\ \onlinecite{Kcnt60}.

Therefore, following
Refs.\ \onlinecite{binderTM82,abinitio2,abinitio1,trio_interactions2}
we added two trio interactions comprising closely spaced
atoms.\cite{ducastelle,trio_interactions2,fiz_met_met}  By trial and
error procedure we were able to achieve a very accurate fit to the six
energies with two trio interactions shown in Fig.\ \ref{fig1} and with
the parameters presented in Table \ref{V}.
\begin{table}
\caption{\label{V}Interactions entering Hamiltonian in
Eq.\ (\ref{ H}) (eV)}
\begin{tabular}{c|c|c|c|c|c}
\hline\hline
$d$&$V^d_1$ & $V^d_2$ & $V^d_3$ & $V_d^{(3)}$ & $E_{ads}$ \\[0.5ex]
\hline
$a$ &-0.1072&-4.119$\cdot10^{-2}$&-7.49$\cdot10^{-2}$&0.242&-1.256\\
$b$ & 0.3427&-0.1084&-0.1268&-2.825$\cdot10^{-2}$& -\\
\hline
\end{tabular}
\end{table}

The pair interactions presented in the table were obtained with the
following parameters of the Morse potential:  $\epsilon = 0.136$ eV,
$r_0 = 5.69$ \AA, and $a= 0.426$.  This can be compared with the
values obtained for the adsorption of potassium on
copper:\cite{morseK/Cu} $\epsilon = 0.466$ eV, $r_0=6$ \AA, and
$a=0.66$.  Taking into account that the systems are very
different, our estimates look reasonable.  A somewhat too small value
of $\epsilon$ which define the attractive interaction between the
potassium atoms may be due to the Coulomb repulsion because of the
considerable and strongly coverage dependent charge transfer between
the potassium and the substrate.\cite{Vasiliev,alkali_review05}  The
value of the adsorption energy in Table \ref{V} also agrees well with
recent {\em ab initio} estimates.\cite{Vasiliev}

Because our statistical approach is based on the grand ensemble, ECIs
in Eq.\ (\ref{ H}) do not depend on the coverage $\rho$ which does not
enter as a parameter in the formalism but is a dependent quantity
calculated according to Eq.\ (\ref{ rho}).  The
concentration-independence may look unphysical because the charge
transfer which strongly influences the Coulomb interatomic interaction
strongly depends on coverage.\cite{alkali_review05} Besides, in a
similar problem in binary alloys it was shown that in the canonical
formalism ECIs do depend on the
concentration.\cite{ducastelle,CEreview}  In
Refs.\ \onlinecite{c_INdep_CE,c_INdep_CE2}, however,  it was shown
that, if properly implemented, both formalisms are equivalent.
Formally in the grand ensemble the concentration independent cluster
interactions (the pair ones, the three-body and higher) cooperate to
reproduce the concentration dependence of the pair interactions of the
canonical formalism.\cite{c_INdep_CE,c_INdep_CE2}

Physically the need for the three-body and higher ECIs can be
understood as follows.  In the case of only pair interactions there
exists a ``particle-hole'' symmetry
\begin{equation} 
\label{ p-h}
H_{\rm pair}=\sum_{ij}V_{i>j}n_in_j-\mu\sum_in_i
=\sum_{i>j}V_{ij}\tilde{n}_i\tilde{n}_j
-\mu^{\prime}\sum_i\tilde{n}_i+C,
\end{equation} 
where $\tilde{n}_i\equiv1-n_i$, $\mu^{\prime}$ is a renormalized
chemical potential, and $C$ a configuration-independent constant.
Because the chemical potential is an adjustable parameter fixing the
coverage, in the case of constant pair interactions $V_{ij}$ it follows
from Eq.\ (\ref{ p-h}) that the free energies calculated for coverage
$\rho$ and $1-\rho$ differ only by the constant $C$.  Thus, the
derivatives of the free energy with respect to $\rho$ whose
singularities correspond to phase transitions (at least, in 2- and 3D
systems) are distributed symmetrically with respect to $\rho=1/2$. This
means that the phase diagram of the system with only pair interactions
is strictly symmetric.\cite{binderTM82} But physically this is rarely
the case, so the presence of higher ECIs is very common.  Quite often
the asymmetry of the diagram is strong which require the presence of
large three-body ECIs comparable in magnitude with the pair
interactions.\cite{binderTM82}  As can be concluded from the value of
the trio interaction $V^{(3)}_a$ in Table \ref{V}, this is also the
case in the potassium adsorbates under consideration.
\section{\label{solution}Potassium adsorption on the (6,0) SWNT}
\begin{figure}
\begin{center}
\includegraphics[viewport =  90 235 462 520, scale = .45]{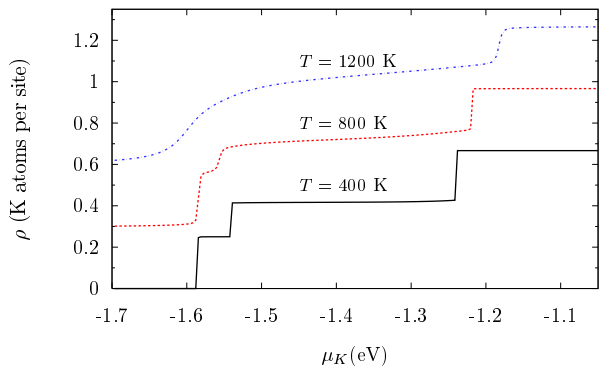}
\end{center}
\caption{\label{fig4}Adsorption isotherms at different temperatures for
the system described by the Hamiltonian in Eq.\ (\ref{ H}) with
parameters given in Table \ref{V}.  Upper curves are shifted up by 0.3 with
respect to the preceding curve for better visibility.}
\end{figure}

The model of potassium adsorption on the (6,0) SWNT
considered in previous section can be solved with the use of the
same TM technique as in Sec.\ \ref{1D} only the TM will be much more
complex than Eq.\ (\ref{ T}).  To account for all interactions shown in
Fig.\ \ref{fig1} we need the TM of a rather large size $2^{14}=16384$,
as explained in Appendix.  Fortunately, only the largest eigenvalue is
needed for our purposes so the efficient technique of finding extremal
eigenvalues of nonsymmetric matrices due to Arnoldi as realized in the
software package {\tt ARPACK}\cite{arpack} could be used.  
Defining the reduced (per site) free energy 
\begin{equation} 
\label{ phi_def}
\phi = -\beta^{-1}\ln\Xi/N,
\end{equation} 
the coverage can be calculated as [see Eq.\ (\ref{ rho0})]
\begin{equation} 
\label{ rho1}
\rho=\sum_{i}\langle n_i\rangle/N= -d\phi/d\mu.
\end{equation} 
The adsorption isotherms for three different temperatures shown in
Fig.\ \ref{fig4} were calculated according to this definition with the
use of the Hellmann-Feynman theorem to improve precision (see
Appendix).  As noted earlier, in the calculations we used Hamiltonian
(\ref{ H}) with parameters from Table \ref{V}.  Though the parameters
fitted the data of Ref.\ \onlinecite{Kcnt60} very accurately, in our
calculations we did not see the quasi-transitions at or close to the
values of the chemical potential shown at Fig.\ 2 of
Ref.\ \onlinecite{Kcnt60}.  Our TM solution, however, is exact up to
the computational errors.  Therefore, to establish the source of the
discrepancy, we took the values of coverages at the lowest temperature
(400 K) curve in Fig.\ \ref{fig2} which to a high accuracy were equal
to 1/4, 5/12, and 2/3 and performed Monte Carlo simulations in the
framework of the canonical ensemble at these coverages.\cite{binder}
The simulations with the use of the Metropolis algorithm were started
at high temperatures and the system was gradually annealed to its
ground state.  At the coverages 1/4 and 2/3 we recovered the structures
of Ref.\ \onlinecite{Kcnt60} while at coverage 5/12 an additional
structure shown in Fig.\ \ref{fig2} was found.  It turned out to have
lower energy than their structure at $\rho = 1/2$.  This is shown in
our Fig.\ \ref{fig3} which is to be compared with Fig.\ 2(a) of
Ref.\ \onlinecite{Kcnt60}.

From the point of view of the CEM, the appearance of a ground state
unaccounted for in {\em ab initio} calculations diminishes the accuracy
of the whole scheme because the ground states are the only ones
directly observable in statistical calculations (as temperature tends
to zero).\cite{CEreview}  Therefore, it is highly desirable that they
entered into the set of the structures calculated {\em ab initio}.
While this point is important for the accuracy of the approach, in the
present paper our main interest is in the universal features of the
thermodynamics which do not depend on the accuracy of the Hamiltonian.
So we believe that as long as the order of magnitude of the
interactions are assessed correctly, the parameters of Table \ref{V}
are sufficiently adequate for our purposes.

Thus, according to our fit the $(2\times2)R0^{\circ}$ structure from
Fig.\ \ref{fig2} is the ground state of Hamiltonian (\ref{ H}) in the
interval of the chemical potentials -1.585 eV $<\mu_K<$ -1.541 eV (see
Fig.\ \ref{fig3}), the 5/12 structure is stable for  -1.541 eV
$<\mu_K<$ -1.243 eV  and the $(\sqrt{3}\times\sqrt{3})R30^{\circ}$
structure with $\rho=2/3$ is the enrgy minimum for $\mu_K$ larger than
-1.243 eV and up to the chemical potential of the bulk potassium
calculated in Ref.\ \onlinecite{Kcnt60} to be equal to -1.15 eV (the
vertical line in our Fig.\ \ref{fig3}).  At finite temperature the
zero-temperature boundaries between the ordered structures give rise to
three quasi-transition steps seen in Fig.\ \ref{fig4}.  For simplicity
we will refer to these transitions in order of their appearance from
left to right as (quasi)transition number one, two, and three,
respectively.
\subsection{Isothermal susceptibility and the universality}
The expression for the susceptibility with respect to the change of the 
chemical potential
\begin{equation} 
\label{ drhodm}
d\rho/d\mu = \beta \sum_i\langle(n_i-\rho)(n_0-\rho)\rangle
\end{equation} 
can be derived form Eqs.\ (\ref{ Xi_def}) and (\ref{ rho0}).  It can be
used to assess the correlation length which we need in Eq.\ (\ref{
eq1}).  As is known, both the susceptibility and the correlation length
diverge at critical points.\cite{wilson_ea72,D1-D4phaseD,nelson_fisher}
Thus, the points of the quasi-phase transitions in 1D at finite
temperature can be identified as the maxima of the correlation length,
as suggested in Ref.\ \onlinecite{porousTM93}.  Below we will determine
in this way the value $\mu_c$ of the critical chemical potential.
Besides, Eq.\ (\ref{ drhodm}) can also be directly related to the
isothermal compressibility because at constant temperature the chemical
potential is proportional to the logarithm of the pressure of the ideal
gas.\cite{HonCNT}  Furthermore, with the use of Eq.\ (\ref{ H-mu}) this
quantity can be directly connected with the magnetic susceptibility of
the Ising model.

\begin{figure}
\begin{center}
\includegraphics[viewport = 150 235 462 520, scale = .45]{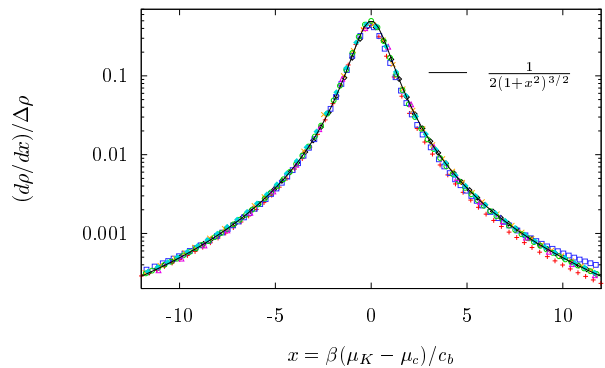}
\end{center}
\caption{\label{fig5}Low temperature behavior of the susceptibility
Eq.\ (\ref{ drhodm}) during three quasi-transitions seen in
Fig.\ \ref{fig4}.  The parameters defining $c_b$ are presented in Table
\ref{transitions}.  $\Delta\rho$ is the change of the coverage during
the transition which is 1/4 in the first and the third transition and
1/6 in the second one.  The data plotted were calculated at
temperatures $T=$ 406 K ($\blacklozenge$), $T=$ 464 K ($\vartriangle$),
and $T=$ 580 K ($\circ$) near the first transition point, at $T=$ 406 K
($\times$) and $T=$ 580 K ($\square$) near the second transition point,
and at $T=$ 464 K ($\lozenge$) and $T=$ 580 K (+) near the third
transition point.}
\end{figure}

As can be seen from Fig.\ \ref{fig4}, the quasi-transitions at the
lowest temperature are so steep that can be easily confounded with the
true first order transitions.  A possible check on whether the
transition is the true one is 
via the susceptibility in Eq.\ (\ref{ drhodm}) which should diverge at
the true phase transition.  We calculated this quantity by numerical
differentiation of $\rho(\mu)$.  The results are plotted in the form of
Eq.\ (\ref{ scaling}) in Fig.\ \ref{fig5}.  The parameters
corresponding to the IPBs are presented in Table \ref{transitions}.  As
can be seen, all values in the table are reasonable from the point of
view of the IPB interpretation: the entropies are all greater than the
lower bound $k_B\ln2\approx0.69k_B$ of the purely 1D model of
Sec.\ \ref{1D} and all enthalpies are notably larger than the
individual interatomic interaction energies in Table \ref{V}.  We,
however, were unable to calculate these values on the basis of an IPB
model.  Qualitatively it is clear that IPBs in our system correspond to
the rearrangement of the atoms from one phase to another and taking
into account the complexity of some of them (see our Fig.\ \ref{fig2}
and Ref.\ \onlinecite{Kcnt60}) the boundary may be not easy to guess.
But even in the simplest case of the quasi-transition 1 between the
empty lattice and the $(2\times2)R0^{\circ}$ phase characterized by six
third-neighbor couplings between the successive atomic layers (see
Fig.\ \ref{fig2}), the na\"{\i}ve calculation $E_b=6V_3^b/2$ [by
analogy with Eq.\ (\ref{ E_b})] gives only 0.38 eV instead of 0.435.
Also the entropy $2.4k_B$ suggests that the interface is rough, as is
usual in 2D Ising-like systems.\cite{abraham76}  In our Monte Carlo
simulations, however, the IPBs were very flat, at least at low
temperatures.  Below we discuss some other possibilities which would
require, however, additional investigations of this issue.
\begin{table}
\caption{\label{transitions}Fitted values of the enthalpy 
and the entropy entering $G_b$ in Eq.\ (\ref{ G_b}) for the three
quasi-transitions seen in Fig.\ \ref{fig4}}
\begin{tabular}{c|c|c}
\hline\hline
Quasi-transition No.&$H_b$ (eV)& $S_b (k_B)$ \\[0.5ex]
\hline
 $1$ &0.435&2.40\\
$2$ & 0.545&4.34\\
$3$ & 0.420&0.80\\
\hline
\end{tabular}
\end{table}
\section{\label{discussion}Discussion}
In the present paper with the use of numerically accurate technique we
were able to resolve the very steep behavior seen in the isotherms of
adsorption on the nanotube surfaces at low
temperatures.\cite{scienceCNT2}  Such behavior is of considerable
interest both from practical and from fundamental points of view.  On
the one hand, it describes the response of the system to small
variations in the external parameters;  on the other hand, the strong
response to the changes of the parameters can provide accurate
information about microscopic interactions.

In our calculations we used a realistic lattice gas model containing
six anisotropic pair interactions and two cluster interactions among
atomic trios derived in the framework of the the cluster expansion
technique\cite{ECI,ducastelle,O/W,CEreview} on the basis of {\em ab
initio} electronic structure calculations.\cite{Kcnt60}  

Despite the complexity of the model, its critical behavior turned out
to be the same as in the 1D Ising (or, equivalently, lattice gas) model
with nearest neighbor interactions.  This agrees with the universality
hypothesis of the theory of critical phenomena yet is a non-trivial
result because contrary to 3D case,\cite{universality72,stanleyRMP} in
1D this hypothesis cannot be justified in the framework of the
renormalization group approach for Hamiltonians with arbitrary
interactions.\cite{prange74} A qualitative explanation may be based on
the very long range correlations present in the system at low
temperature.  The correlation length which can be assessed from the
right hand side of Eq.\ (\ref{ drhodm}) is of $O(1/c_b)$ and reaches
values of $O(10^4)$ as can be estimated from Table \ref{transitions}.
This means that the structures are correlated at very long distances
and, using the language of the renormalization group and the Ising
model, the block spin transformation can be efficient in theoretical
description of the system.  Because the tube diameter is much smaller
than the correlation length, the block spins will comprise all the
spins around the tube circumference as well as considerable block of
sites along the tube.  In this picture the interactions of sufficiently
short range will connect only the nearest neighbor block spins thus
making the system effectively equivalent to strictly 1D model with only
nearest neighbor interactions.

The correlation length in $O(10^4)$ of lattice spacings along the tube
means that the whole nanotube can be covered by the ordered structure
at temperatures as high as 400 K.  (We note that only in infinite
systems the long range order should be broken in 1D;\cite{ll} in a
finite system the order can extend along the whole tube
length.\cite{1Dorder})  Thus, from the hydrogen storage standpoint, at
ambient conditions the potassium structure can be treated as an inert
(in statistical sense) substrate while the hydrogen molecules treated
within a statistical approach which can be based on the formalism
developed in the present paper.

In this study we concentrated on the universality for the following
reasons.  First, because we had at our disposition the energies of only
six {\em ab initio} calculated structures, the accuracy of the cluster
Hamiltonian was rather poor.  Therefore, only orders of magnitude of
the quantities of interest could be calculated judging from the fact
that even 60 structures calculated in Ref.\ \onlinecite{O/W} did not
allow to calculate the phase transition temperatures with accuracy
better than 50\%.  The universal behavior, however, is the same for all
Hamiltonians belonging to the same universality class.  The second
reason was that in many cases the surface structures are not
commensurate with the substrate.\cite{scienceCNT2,referee1PRB}  Yet
they can be as good hydrogen adsorbers as the commensurate structures.
But, as we noted in the Introduction, the lattice structure is not
needed for the critical behavior to be universal, as the gas-liquid
transitions in 1-, 2-, and 3D systems
show.\cite{wilson_ea72,1Dliquid86,2Dliquid92}  According to
Ref.\ \onlinecite{rivier} the liquid can be viewed as a crystalline
state filled with topological defects, such as dislocations and
disclinations.  The same can be said about the incommensurate surface
layers.\cite{AgAg}  Therefore, one might expect that the universal
critical behavior may take place also in the incommensurate cases.
This prediction should be amenable to experimental verification on the
isotherms of SWNTs covered with incommensurate phases of inert
gases.\cite{scienceCNT2}

The third reason for studying the universality was that while the TM
technique is an accurate and efficient tool for treating the ordering
of coherent adsorbates on surfaces of small nanotubes, the size of TM
grows exponentially with the tube diameter and with the range of
interactions.  This means that for only slightly larger tube or
longer-ranged interaction the TM will became unmanageably large.  There
exist viable alternatives to the TM in solution of this kind of
problems: the mean field approximation and especially the Monte Carlo
method.\cite{binderTM82,porousTM93,pore_MF02,pore_monte_carlo04} Both
techniques, however, meet with difficulties in treating the fine
details of the abrupt phase transition-like changes seen on the
adsorption isotherms.  The results obtained in the present paper are
aimed at resolving this difficulty.  The mean field or the Monte Carlo
methods can accurately predict the position of the transition while the
universality in the transition curves observed in our study should
provide its fine details.  There remains the problem of finding the
values of the two parameters which describe the behavior
quantitatively.  A block-spin renormalization group and/or the
low-temperature expansion are probable candidate tools for attacking
this problem.  Further work is needed to clarify this point.
\begin{acknowledgments}
The authors acknowledge CNRS for support of their collaboration.  One
of the authors (V.I.T.) expresses his gratitude to Universit{\'e}
de Strasbourg and IPCMS for their hospitality.
\end{acknowledgments}
\appendix*
\section{\label{TM}Sparse transfer matrices}
Our TM approach belongs to the general category of TM methods based on
sparse matrices initiated in
Refs.\ \onlinecite{screwed_cilinder,screwed_cilinder2}; further
bibliography can be found in Ref.\ \onlinecite{sparse85}.  The problem
of adsorption on triangular lattice with account of the second neighbor
and trio interactions was previously studied within similar framework
in Ref.\ \onlinecite{binderTM82} but no details were given.  We believe
that our technique presented below is particularly simple and easy to
use.

The advantage of using sparse TMs is that instead of $l^2$ matrix
elements of a conventional dense $l\times l$ TM matrix (see, e.\ g.,
Ref.\ \onlinecite{ducastelle}) one deals with matrices containing only
$O(l)$ nontrivial entries.  Because the size of TM scales with the
range of interactions $R$ exponentially as
\begin{equation} 
\label{ l}
l=2^R
\end{equation} 
and in practical calculations reaches significant values (e.\ g.,
$2^{14}=16384$ in the present study), the gain in numerical efficiency
from using sparse TMs can be enormous.

The interaction range $R$ in Eq.\ (\ref{ l}) for Hamiltonian (\ref{ H})
is defined as the longest range of the cluster interactions it
contains.  The range of a cluster interaction $V^{(m)}_{i_1\dots
i_n}n_{i_1}\dots n_{i_m}$ is defined as
\begin{equation} 
\label{ R}
R=i_{max}-i_{min}, 
\end{equation} 
where $i_{max}$ and $i_{min}$ are the maximal and the minimal indices
amonf $i_1,\dots,i_m$.  For example, the cluster interaction
$n_in_{i+1}n_{i+2}$ has the range $R=2$.

The finite range of interactions in the Hamiltonian makes possible a
recursive calculation of the partition function.  This is because when
adding a site to the system consisting of $K\geq R$ sites only the
interactions with the last $R$ sites need be taken into account.  The
accounting can be done with the use of the vector partition function
$\vec{Z}^{(K)}$ whose components are the partial traces over all except
the last $R$ sites (the sites are numbered from right to left)
\begin{equation} 
\label{ Z_partial}
Z^{(K)}_{n_K,n_{K-1},\dots,n_{K-R+1}}
=\mbox{Tr}_{n_1,n_2,\dots,n_{K-R}} \exp(-H^{(K)}),   
\end{equation} 
which can be visualized as
\begin{equation} 
\label{ trace}
Z^{(K)}_{\underbrace{{\bullet\circ\dots\bullet}}_R}
=\overbrace{{\underbrace{{\bullet\circ\dots\bullet}}_R}\;
{\ast\ast\dots\ast\ast}}^K,
\end{equation} 
where the empty and filled circles correspond to the empty ($n_i=0$) or 
filled ($n_i=1$) sites in Eq.\ (\ref{ Z_partial})
while asterisks denote the sites over which the trace over the two
possible values of filling has been taken;  $H^{(K)}$ is
Hamiltonian (\ref{ H}) for a $K$-site system.
The partition function is found from (\ref{ Z_partial}) as
\begin{equation} 
\label{ Z_K}
Z^{(K)}=\sum_{\bar{\alpha}={\bar{0}}}^{\overline{2^R-1}}
Z^{(K)}_{\bar{\alpha}}.
\end{equation}
Here the bar over the number denotes that its binary representation is
meant 
\begin{equation} 
\label{ barC}
\bar{A}=(a_{R-1}\dots a_1a_0)_R, 
\end{equation} 
where $a_k=0,1$ correspond to the filling of site $k$.  The subscript
$R$ reminds that the term within parentheses is the binary
representation, not the product, and that its length is equal to $R$.
For example, $\bar{1}=00\dots01$ with $R-1$ zeros before the unity 
means that there is $R-1$ empty sites before the filled one.

The general form of the TM can be understood from the recurrence
equation
\begin{widetext}
\begin{equation} 
\label{ Xi}
\left(\!\!\!
\begin{array}{c}
\circ\circ\dots\circ\circ\\
\underbrace{\circ\circ\dots\circ\bullet}_R\\
\vdots\\[-0.5ex]
\circ\bullet\dots\bullet\bullet\\
\bullet\circ\dots\circ\circ\\
\bullet\circ\dots\circ\bullet\\[-0.5ex]
\vdots\\[-0.5ex]
\bullet\bullet\dots\bullet\bullet
\end{array}
\!\!\right)^{(N)}\!\!\!=
\left(
\begin{array}{cccccccc}
1&1&0&0&\dots&0&0&0\\
0&0&1&1&\dots&0&0&0\\
\vdots&\vdots&\vdots&\vdots&\ddots&\vdots&\vdots&\vdots\\ 
0&0&0&0&\dots&0&1&1\\
b_0&b_1&0&0&\dots&0&0&0\\
0&0&b_2&b_3&\dots&0&0&0\\
\vdots&\vdots&\vdots&\vdots&\ddots&\vdots&\vdots&\vdots\\ 
0&0&0&0&\dots&0&b_{2^R-2}&b_{2^R-1}
\end{array}
\right)_{(N)}
\!\left(\!\!\!
\begin{array}{c}
\circ\circ\dots\circ\circ\\
\underbrace{\circ\circ\dots\circ\bullet}_R\\
\vdots\\[-0.5ex]
\circ\bullet\dots\bullet\bullet\\
\bullet\circ\dots\circ\circ\\
\bullet\circ\dots\circ\bullet\\[-0.5ex]
\vdots\\[-0.5ex]
\bullet\bullet\dots\bullet\bullet
\end{array}
\!\!\right)^{(N-1)}
\end{equation} 
\end{widetext}
where the column vectors correspond to $\vec{Z}^{(N-1)}$ and
$\vec{Z}^{(N)}$ with the components denoted by their subscripts in
Eq.\ (\ref{ trace}) for brevity.  The subscript $N$ of the TM is the
site index for all $b_{\bar{\alpha}}$ entering the matrix.  We note
that we use the same symbol $N$ for the system size and for the
recurrent relation in order to stress that at every iteration we obtain
the (vector) partition function of a system of size $N$, i.\ e., that
the partition functions obtained at intermediary steps are not in any
way deficient.

The structure of TM in (\ref{ Xi}) is physically transparent. Having
added site $N$ to the system consisting of $N-1$ sites we first have to
account for the interaction of this site with the rest of the system
and then take the trace over the ($N-R$)-th site because with the
radius of interactions being $R$ all interactions of this site with the
rest of the system have already been taken into account.  Taking the
trace amounts to adding with appropriate weights two $Z^{(N-1)}$
differing by the filling of site $N-R$.  In the case when site $N$ is
empty the weights are equal to unity because the empty site does not
interact with anything and the interaction energy is zero.  These terms
occupy the upper half of the TM (\ref{ Xi}).  The lower half of the
matrix contains the terms corresponding to the interaction of the {\em
occupied} site $N$ with the rest of the system.  The term
\begin{equation} 
\label{ b_i}
b_{\bar{\alpha}N}=\exp(-\beta\Delta E_{\bar{\alpha}N}) 
\end{equation} 
is the Boltzmann weight corresponding to the interaction of the atom at
site $N$  with the configuration of atoms corresponding to
$Z^{(N-1)}_{\bar{\alpha}}$; $\Delta E_{\bar{\alpha}N}$ in (\ref{ b_i})
is the energy of interaction of the atom at site $N$ with configuration
$\bar{\alpha}$ on sites $N-1$, $N-2$, $\dots$, $N-R$.
\subsection{Application to adsorption on (6,0) nanotube} 
\begin{figure}
\begin{center}
\includegraphics[viewport = 50 700 496 850, scale = .55]{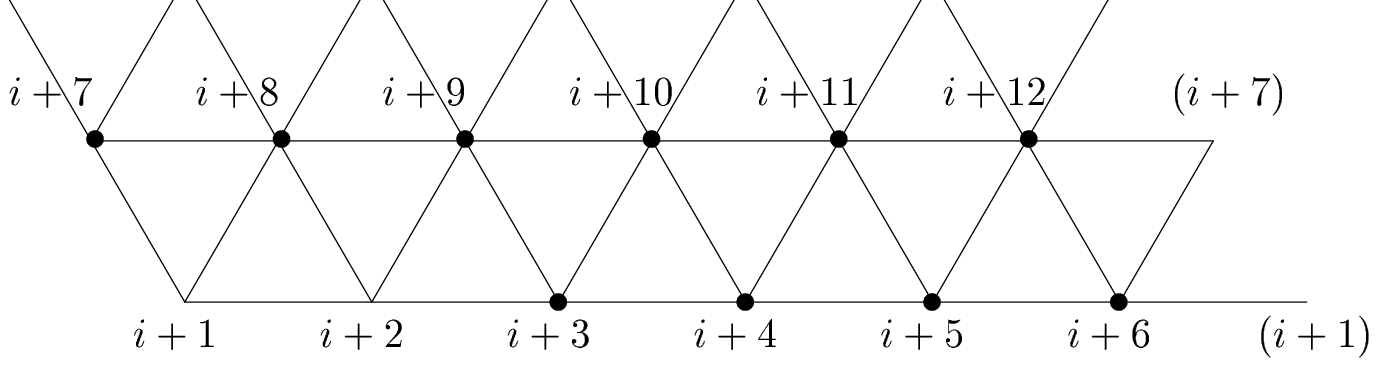}
\end{center}
\caption{\label{fig6}Enumeration of sites on the (6,0) nanotube used in
the construction of the TM.  Black dots denote the fourteen neighbors
of site $i+17$.  The 14-th neighbor $i+3$ is defined by the interaction
$V_3^b$ in Fig.\ \ref{fig1}.}
\end{figure}

In
Fig.\ \ref{fig6} are shown both the enumeration of sites we chose for
the (6,0) nanotube and the fourteen sites along the path with which
atom at site $i+17$ can interact if they are also filled with atoms.
The furthest neighbor site $i+3$ is defined by the the 3rd neighbor
interaction $V_3^b$ which can reach it (see Fig.\ \ref{fig1}).  Thus,
according to Eq.\ (\ref{ R}) $R=14$ and the size of our TMs is
$2^{14}=16384$.  This is the number of configurations we need to
account for in our transfer matrices.
From Fig.\ \ref{fig6} one can see that as the sites are being added one
after another in the top row the relative placement of the 14th
neighbor change with respect to the added site.  Because of this the
transfer matrices for neighbor sites are different, except for sites
$i+15$ and $i+16$ which is due to the particular interactions entering
our Hamiltonian.  This can be seen from Table \ref{3} where the pair
interactions accounted for in the Boltzmann factors $b_k$ entering the
TMs are presented.  Similar tables can be composed for the trio
interactions.
\begin{table}[t]
\caption{\label{3}Interactions of atoms in the top row on
Fig.\ \ref{fig6} with atoms on fourteen preceding sites (index $i$ has
been omitted for brevity).  The arrows point to the value they
represent.}
\begin{tabular}{c|c|c|c|c|c}
\hline\hline
Neighbor No. & 13 & 14 & 15-16 & 17 & 18 \\[0.5ex]
\hline
1 & $V_1^b$ &  $V_1^a$ & $\leftarrow$ & $\leftarrow$ & $\leftarrow$ \\
2 & $V_2^b$ & $\leftarrow$ & $V_3^a$ &$\leftarrow$ &$\leftarrow$ \\
4 & 0 & $\leftarrow$ & $\leftarrow$ & $V_3^a$ & $\leftarrow$\\
5 & $V_2^b$ & $\leftarrow$ & $\leftarrow$ & $\leftarrow$ & $V_1^a$\\
6 & $V_1^b$ & $\leftarrow$ & $\leftarrow$ & $\leftarrow$ &
$\leftarrow$ \\
7 & $V_2^a$ & $V_1^b$ & $\leftarrow$ & $\leftarrow$ & $\leftarrow$ \\
8 & $V_3^b$ & $\leftarrow$ & $V_2^b$ & $\leftarrow$ & $\leftarrow$ \\
11 & 0 &  $\leftarrow$ & $\leftarrow$ & $\leftarrow$ & $V_2^b$\\
12 & $V_3^b$ & $\leftarrow$ & $\leftarrow$ & $\leftarrow$ &
$\leftarrow$ \\
13 & 0 & $V_2^a$ & $\leftarrow$ & $\leftarrow$ & $\leftarrow$ \\
14 & 0 & $\leftarrow$ & $V_3^b$ & $\leftarrow$ & $\leftarrow$ \\
\hline
\end{tabular}
\end{table}

It is easy to see that the structure of the TMs repeats after each six
steps, for example, when the row gets filled.  This allows one to
compute the reduced free energy of the system Eq.\ (\ref{ phi}) through
the logarithm of the largest eigenvalue $\lambda_+$ of the product of
six TMs, e.\ g., of those corresponding to sites from $i+13$ to $i+18$
in Fig.\ \ref{fig6} as
\begin{equation} 
\label{ phi6}
\phi = -k_BT\ln\lambda_+/6.
\end{equation} 
\subsection{Adsorption isotherms}
To draw the adsorption isotherm one has to calculate the derivative of
$\phi$ with respect to the chemical potential [see Eq.\ (\ref{
rho1})].  With the fast variation of the derivative in the most
interesting region in the vicinity of the quasi-transition, the
numerical differentiation can be unreliable.  More accurate results can
be obtained with the use of the Hellmann-Feynman theorem:
\begin{equation} 
\label{ H-F}
\rho = -\frac{d\phi}{d\mu} = \frac{1}{6}\frac{\langle+|d\hat{M}_6
/d(\beta\mu)|+\rangle}
{\langle+|\hat{M}_6|+\rangle},
\end{equation} 
where $\hat{M}_6$ is the product of the six TMs, as explained above,
$|+\rangle$ is the eigenvector corresponding to $\lambda_+$, and
$\langle+|$ the eigenvector of the transposed matrix because our TMs
are not symmetric.  Due to the simplicity of our TMs the derivative in
Eq.\ (\ref{ H-F}) is very easy to calculate: the upper part of each of
the six TMs entering $\hat{M}_6$ should simply be successively set to
zero while the lower part remains the same because the differentiation
does not change the exponential function
$\exp(\beta\mu)$.

\end{document}